\documentclass{iopart}

\usepackage{graphicx}
\usepackage{url}
\expandafter\let\csname equation*\endcsname\relax
\expandafter\let\csname endequation*\endcsname\relax
\usepackage{amsmath}
\usepackage{array}
\usepackage{amssymb}
\newcommand{\real} {\mathop{\mathrm{Re}}\nolimits}

\begin{document}

\title[Gyrotropic-nihility state] {Gyrotropic-nihility state in \\ a composite ferrite-semiconductor structure}

\author{Vladimir R Tuz$^{1,2}$}

\address{$^1$ Institute of Radio Astronomy of National Academy of Sciences of Ukraine, 4, Chervonopraporna
Street, Kharkiv 61002, Ukraine}
\address{$^2$ School of Radio Physics, Karazin Kharkiv National University, 4,
Svobody Square, Kharkiv 61022, Ukraine}
\ead{tvr@rian.kharkov.ua}

\begin{abstract}
Characteristics of the gyrotropic-nihility state are studied in a
finely-stratified ferrite-semiconductor structure, which is under an
action of an external static magnetic field. Investigations are
carried out with the assistance of the effective medium theory,
according to which the studied structure is approximated as a
uniform gyroelectromagnetic medium. The theory of the
gyrotropic-nihility state is developed in terms of the eigenwaves
propagation in such gyroelectromagnetic medium. The frequency and
angular dependencies of the transmittance, reflectance and
absorption coefficient are presented. It turns out that in the
frequency band around the frequency of gyrotropic-nihility state the
studied structure appears to be matched to free space with both the
refractive index and the wave impedance which results in its high
transmittance almost in the entire range of angles of the
electromagnetic wave incidence.
\end{abstract}

\submitto{\JPCM}
\maketitle

\section{Introduction}

The conception of ``nihility'' media
\cite{Lakhtakia_IntIRMMWaves_2002}  was firstly introduced into
electromagnetics in regard to the theory of metamaterials. It
determines a distinctive exotic state of a hypothetical lossless
medium whose material parameters happen to be zero quantities. It
corresponds to the fulfilment of the constitutive relations where
the electric and magnetic flux densities are both equal to zero,
$\vec D = 0$ and $\vec B = 0$. Thereby, the nihility medium exhibits
a zero refractive index, so, there are $\nabla \times \vec E = 0$
and $\nabla \times \vec H = 0$, and the wave propagation in such
extreme-parameter medium becomes forbidden in the absence of sources
therein. It means that the directionality of the phase velocity in
relation to the wavevector is a non-issue for this medium. Although
the nihility media determined in such a way are unachievable in
nature, they can be approximately simulated at a specified
frequency.

As such approximation the theory of ``epsilon-near-zero'' and
``mu-near-zero'' materials \cite{Ziolkowski_PhysRevE_2004,
Alu_PhysRevB_2007, Davoyan_OptExpress_2013} can be considered since
it is fundamentally identical with the conception of nihility media.
In this theory the Drude medium model is consistently used to
simulate a certain dispersive medium in which real parts of both
complex permittivity and complex permeability simultaneously acquire
zero at a specified frequency, and the refractive index of the
resulting medium appears to be zero value, too. It means that while
the zero refractive index is not matched to free space, the wave
impedance is matched one. The most intriguing properties of such
medium is the nearly zero phase progression of propagating waves,
which may be of interest in the problems of transformation optics
\cite{Wei_JOpt_2011}, and can be employed for the efficient energy
tunneling \cite{Liu_PhysRevLett_2008} and for the design of the next
generation of waveguides \cite{Edwards_PhysRevLett_2008}.
Furthermore, a strong material dispersion leads to possibility of
medium properties changing  form being opaque ``metals''  to
transparent dielectrics with a small variation of the material
parameters around the epsilon-near-zero and mu-near-zero states.

However, further mixing together electric and magnetic responses of
the medium, for example, by utilizing magneto-optic (gyrotropic)
and/or bi-anisotropic effects, allows one to drastically change the
peculiarities of the wave propagation in nihility media. For
instance, in \cite{Tretyakov_JEMWA_2003} it has been proposed a
possible way to achieve a nihility state in a bi-isotropic medium
using canonical chiral wire particles. The dispersive
characteristics of the effective constitutive parameters of the
medium are calculated on the basis of the Maxwell-Garnett mixing
rule. It is found that, at a specified frequency, the real parts of
both complex effective permittivity and complex effective
permeability happen to be close to zero while the chirality
parameter is maintained at a finite value. This distinct
extreme-parameter state is referred to as ``chiral-nihility'', and
it turns out that, in contrast to the conventional nihility medium,
in the chiral-nihility both the refractive index and the wave
impedance appear to be matched to free space, and the directionality
of the phase velocity relative to the wavevector would not be a
non-issue any more. So, the wave propagation is allowed in
chiral-nihility media, and, in particular, this wave propagation can
be described in the term of two orthogonal circularly polarized
eigenstates. It means that in this medium there are two eigenwaves
with right- and left-circularly polarized states, and, noteworthy,
one of these eigenwaves experiences a backward
propagation\footnote{We consider a backward wave as a wave in which
the direction of the Poynting vector is opposite to that of its
phase velocity. We assume that the direction of the energy flux is
coincident with the direction of exponential decay of the wave's
field, caused by dissipation and absorption of the wave energy in
the medium \cite{Lindell_MOP_2001, Shevchenko_PhysUsp_2007}.}. It
results in some exotic characteristics (e.g. wave tunneling and
rejection) in the waves interaction with a single layer and a
multilayer system consisting of such chiral-nihility media
\cite{Qiu_JOSAA_2008, Xiangxiang_IEEEAPMagazine_2009, Tuz_PIER_2010,
Gevorgyan_JOpt_2013}.

More recently, the extreme-parameter nihility state has been also
found for the other fundamental class of reciprocal bi-anisotropic
media, namely, for omega materials \cite{Tretyakov_Metamat_2012}.
The omega material is a reciprocal bi-anisotropic medium
characterized with an antisymmetric magnetoelectric dyad (e.g., it
is a composite material formed by inserting conductive
$\Omega$-shaped particles into a homogeneous dielectric host). In
contrast to chiral media where the chirality breaks the symmetry of
the propagation constants of the circularly polarized eigenwaves
while the wave impedances are not affected, in omega materials the
properties are dual: the magneto-electric coupling breaks the
symmetry of the wave impedances while the propagation constants
remain symmetric. In this case, too, the extreme ``omega-nihility''
state of the omega material appears at a specified frequency, when
real parts of both complex permittivity and complex permeability of
the medium happen to be close to zero, and the magnetoelectric
parameter alone defines the material response. Among other effects,
the omega-nihility material provides an extreme asymmetry in
reflection from a material slab: the reflection coefficients from
the two opposite sides differ by sign, while the transmission
coefficient is symmetric as in any conventional reciprocal material
slab.

It is a common knowledge that, besides chiral media, the circularly
polarized eigenstates are also inherent to circularly birefringent
(gyrotropic) materials (e.g. plasmas, ferrites and semiconductors)
in the presence of an external static electric or magnetic field,
when this field is biased to the specimen in the longitudinal
geometry relative to the direction of the wave propagation (Faraday
configuration). Such gyrotropic media are characterized by the
permittivity or permeability tensor with non-zero off-diagonal
elements (gyrotropic parameters). It is no wonder, that properly
combining together gyroelectric and gyromagnetic materials into a
uniform gyroelectromagnetic structure \cite{Prati_JEMWA_2003} allows
one to reach a ``gyrotropic-nihility'' state at a specified
frequency. In particular, as it has been shown in
\cite{Tuz_PIERB_2012}, in a composite finely-stratified
ferrite-semiconductor structure the conditions of
gyrotropic-nihility state are valid in the microwave band near the
frequencies of ferromagnetic and plasma resonances. In this case
real parts of diagonal elements of both complex effective
permittivity and complex effective permeability tensors of such
artificial medium simultaneously acquire zero, while the
off-diagonal elements appear to be non-zero quantities. It is
revealed that in this structure, in the certain frequency band, the
backward propagation takes a place for one of the circularly
polarized eigenwaves which can lead to some unusual features of the
system and provides an enhanced polarization rotation, impedance
matching and complete light transmission.

The objective of this paper is a formalization of the theory of the
gyrotropic-nihility state in terms of the eigenwaves propagation in
such composite finely-stratified ferrite-semiconductor structure.

\section{Constituents of a composite ferrite-semiconductor structure}

The essence of any nihility medium is extreme characteristics of its
constitutive parameters. These extreme characteristics appear in the
small region near singular points of dispersion curves where real
parts of both complex permittivity and complex permeability
simultaneously make transitions from negative to positive values or
vice versa. It is worth noting that media for which these
transitions separately exist for permittivity or permeability may be
directly found in nature. A well-known example is an electron gas,
in which, due to the conduction current created by the drift of free
electric charges, may efficiently interact with radiation as a
continuous medium characterized by a Drude dispersion model, near
the plasma frequency the real part of permittivity appears to be
close to zero \cite{Akhiezer_book_1975}. At infrared and optical
frequencies some low loss noble metals (e.g. silver, gold),
semiconductors (e.g. indium antimonide), and some polar dielectrics
(e.g. silicon carbide) may behave an epsilon-near-zero state near
their plasmas frequencies. At the same time, a mu-near-zero state is
inherent to magnetic materials (e.g. ferromagnets, ferrimagnets) in
the vicinity of a gyromagnetic resonance \cite{Duncan_ProcIRE_1957}.
Thereby, although any nihility medium does not exist in nature in
its pure form, there is a possibility of obtaining nihility
artificially by mixing together materials which manifest
epsilon-near-zero and mu-near-zero states in the same frequency
range.

In particular, such an opportunity exists in the microwave part of
spectrum where a low temperature magnetized plasma can be obtained
whose dispersion properties are simultaneously defined by tensors of
effective permittivity $\hat \varepsilon_{eff}$ and effective
permeability $\hat \mu_{eff}$. Its constituents are micron ferrite
grains (e.g. yttrium iron garnet, YIG) admixed to the magnetized
electron-ion plasma \cite{Rapoport_PhysPlasmas_2010}. The dispersion
of permeability is caused by the high frequency magnetization of the
grain subsystem and is important in the vicinity of the frequency of
a ferromagnetic resonance, which coincides with the electron
cyclotron frequency. Since the resulting medium is under an action
of the static magnetic field, an effect of gyrotropy appears. We
predict that through careful adjustment of the ratio of the magnetic
grain fraction in the plasma one can reach epsilon-near-zero and
mu-near-zero states simultaneously at the same frequency, and, thus,
achieve a gyrotropic-nihility.

As an alternative, a layered heterostructure can be considered
\cite{Tuz_PIERB_2012, Shramkova_PIERM_2009, Wu_JPhysCondMatt_2007}.
It is constructed by periodically arranging into a certain uniform
system of gyroelectric (semiconductor) and gyromagnetic (ferrite)
layers, provided that these layers are optically thin and their
underlying materials correspondingly have epsilon-near-zero and
mu-near-zero states in the same frequency range. If the constitutive
layers as well as the period of the final structure are optically
thin (i.e. it is a finely-stratified one) the effective medium
theory can be consistently applied in order to identify its
homogenized material parameters. It results in the consideration of
the finely-stratified structure as a homogeneous gyroelectromagnetic
medium described by tensors of effective permittivity $\hat
\varepsilon_{eff}$ and effective permeability $\hat \mu_{eff}$ which
possess some dispersion characteristics.

\begin{figure}[htbp]
\centerline{\includegraphics[width=0.7\columnwidth]{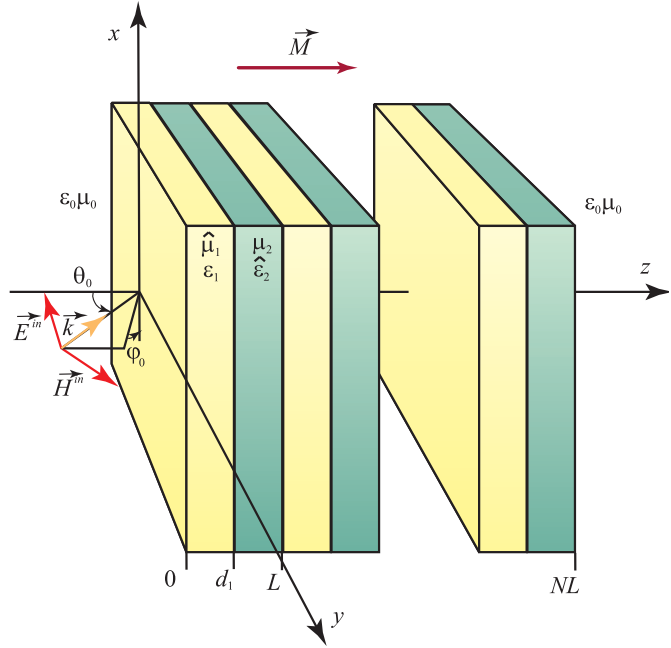}}
\caption{(Color online) A composite finely-stratified
ferrite-semiconductor structure, which is under an action of an
external static magnetic field.} \label{fig:fig1}
\end{figure}

Thereby, further in this paper we consider a stack of $N$ identical
double-layer slabs (unit cells) periodically arranged along the
$z$-axis (figure~\ref{fig:fig1}). Each unit cell is constructed by
juxtaposition together of ferrite (with constitutive parameters
$\varepsilon_1$, $\hat \mu_1$) and semiconductor (with constitutive
parameters $\hat \varepsilon_2$, $\mu_2$) layers with thicknesses
$d_1$ and $d_2$, respectively. The structure's period is $L = d_1 +
d_2$, and along the $x$ and $y$ directions the system is infinite.
We suppose that the structure is a finely-stratified one, i.e. its
characteristic dimensions $d_1$, $d_2$ and $L$ are all much smaller
than the wavelength in the corresponding layer $d_1\ll \lambda$,
$d_2 \ll \lambda$, and period $L \ll \lambda$  (the long-wavelength
limit). An external static magnetic field $\vec M$ is directed along
the $z$-axis. The input $z \le 0$ and output $z \ge NL$ half-spaces
are homogeneous, isotropic and have constitutive parameters
$\varepsilon_0$, $\mu_0$.

We use common expressions for underlying constitutive parameters of
normally magnetized ferrite and semiconductor layers with taking
into account the losses. For ferrite layers the permittivity and
permeability are defined in the form \cite{Gurevich_book_1963,
Collin_book_1992}:
\begin{equation}
\varepsilon_1  = \varepsilon_f,~~~~~\hat\mu_1=\left( {\begin{matrix}
   {\mu_1^T } & {\rmi\alpha} & 0  \cr
   {-\rmi\alpha } & {\mu_1^T } & 0  \cr
   0 & 0 & {\mu_1^L }  \cr
 \end{matrix}
} \right), \label{eq:ferrite}
\end{equation}
where $\mu_1^T=1+\chi' + \rmi\chi''$,
$\quad\chi'=\omega_0\omega_m[\omega^2_0-\omega^2(1-b^2)]D^{-1}$,
$\chi''=\omega\omega_m b[\omega^2_0+\omega^2(1+b^2)]D^{-1}$,
$\quad\alpha=\Omega'+\rmi\Omega''$,
$\quad\Omega'=\omega\omega_m[\omega^2_0-\omega^2(1+b^2)]D^{-1}$,
$\Omega''=2\omega^2\omega_0\omega_m bD^{-1}$, $\quad
D=[\omega^2_0-\omega^2(1+b^2)]^2+4\omega^2_0\omega^2 b^2$,
$\omega_0$ is the Larmor frequency and $b$ is a dimensionless
damping constant.

For semiconductor layers the permittivity and permeability are
defined as follows \cite{Bass_book_1997}:
\begin{equation}
\hat\varepsilon_2=\left( {\begin{matrix}
   {\varepsilon_2^T} & {\rmi\beta} & 0 \cr
   {-\rmi\beta } & {\varepsilon_2^T } & 0 \cr
   0 & 0 & {\varepsilon_2^L } \cr
\end{matrix}
} \right),~~~~~\mu_2=\mu_s, \label{eq:semicond}
\end{equation}
where $\varepsilon_2^T=\varepsilon_l\left[ {1-\omega_p^2
(\omega+\rmi\nu)[\omega((\omega+\rmi\nu)^2-\omega_c^2)]^{-1}}\right]$,
$\varepsilon_2^L=\varepsilon_l\left[{1-\omega_p^2[\omega(\omega+\rmi\nu)]^{-1}}\right]$,
$\beta=\varepsilon_l\omega_p^2\omega_c[\omega((\omega+\rmi\nu)^2-\omega_c^2)]^{-1}$,
$\varepsilon_l$ is the part of permittivity attributed to the
lattice, $\omega_p$ is the plasma frequency, $\omega_c$ is the
cyclotron frequency and $\nu$ is the electron collision frequency in
plasma.

\begin{figure}[htbp]
\centerline{\includegraphics[width=1.0\columnwidth]{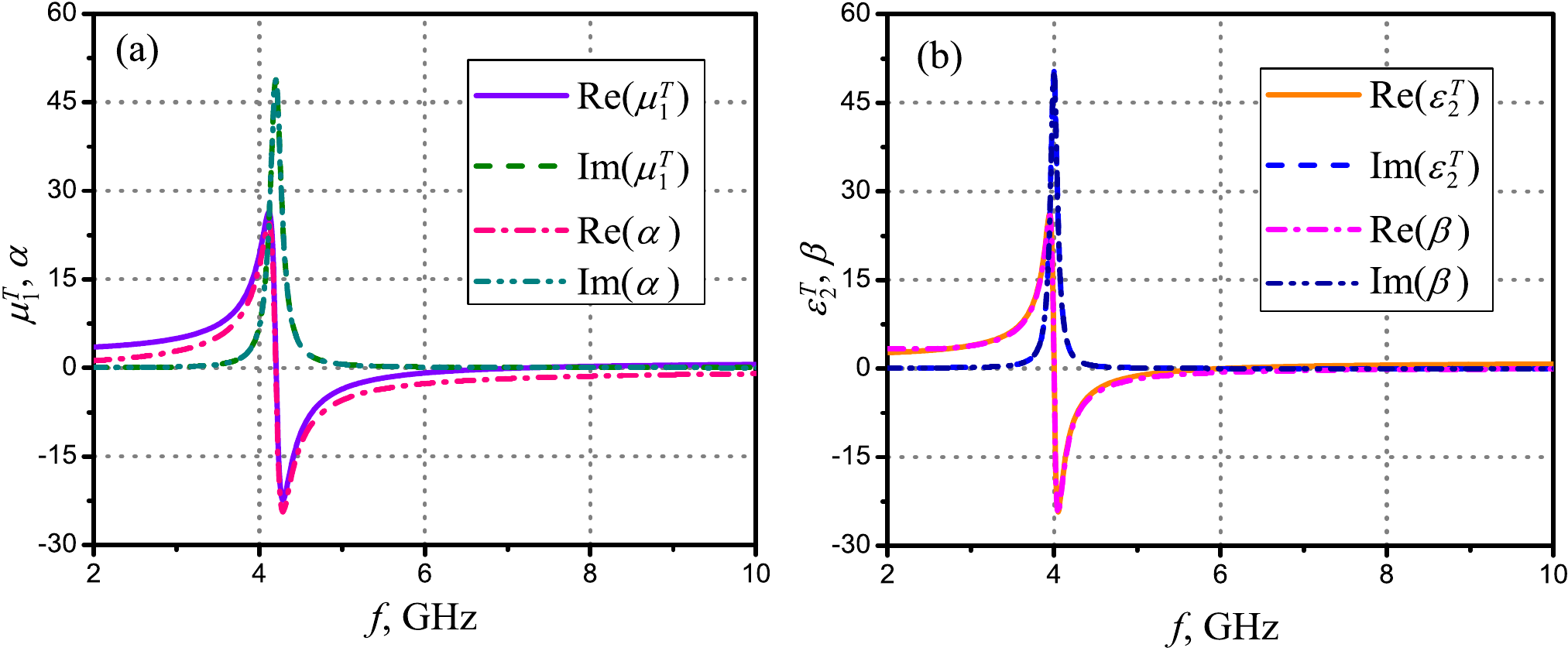}}
\caption{Frequency dependencies of (a) permeability and (b)
permittivity of ferrite and semiconductor layers, respectively. We
use typical parameters for these materials in the microwave region.
For the ferrite layers, under saturation magnetization of 2000~G,
parameters are: $\omega_0/2\pi=4.2$~GHz, $\omega_m/2\pi=8.2$~GHz,
$b=0.02$, $\varepsilon_f=5.5$. For the semiconductor layers,
parameters are: $\omega_p/2\pi=4.5$~GHz, $\omega_c/2\pi=4.0$~GHz,
$\nu/2\pi=0.05$~GHz, $\varepsilon_l=1.0$, $\mu_s=1.0$.}
\label{fig:fig2}
\end{figure}

The frequency dependencies of the real and imaginary parts of the
underlying permeability and permittivity calculated using
equations~(\ref{eq:ferrite}) and (\ref{eq:semicond}) are presented
in figure~\ref{fig:fig2}. Note that imaginary parts of $\mu^T_1$,
$\alpha$ and $\varepsilon^T_2$, $\beta$ are so close to each other
that the curves of their frequency dependencies coincide in the
corresponding figures.

\section{Homogenized material parameters of a gyroelectromagnetic medium}

In the long-wavelength limit, in order to reveal the
gyrotropic-nihility state conditions which are attainable in the
studied structure, the effective medium theory can be engaged
\cite{Tuz_PIERB_2012, Lakhtakia_Optic_2003}. From the viewpoint of
this theory, the periodic structure is approximately represented as
an anisotropic (gyroelectromagnetic) uniform medium whose optical
axis is directed along the structure periodicity, and this medium is
described with some tensors of effective permittivity
$\hat\varepsilon_{eff}$ and effective permeability $\hat\mu_{eff}$
which should be retrieved.

Let us consider a unit cell of the studied structure. It is made of
two layers $0<z<d_1$ and $d_1<z<L$ of dissimilar materials whose
constitutive relations are as follows:
\begin{equation}
\left.\begin{array}{c} \vec D=\varepsilon_1\vec E \\
                       \vec B=\hat\mu_1\vec H \end{array}\right\}~~0<z<d_1,
~~~~~
\left.\begin{array}{c} \vec D=\hat\varepsilon_2\vec E \\
                       \vec B=\mu_2\vec H \end{array}\right\}~~d_1<z<L.
\label{eq:constitutive}
\end{equation}
In general form, in Cartesian coordinates, the system of Maxwell's
equations for each layer has a form\footnote{Hence, the field
vectors will be supposed to contain a time factor of the form
$\exp(-\rmi \omega t)$.}
\begin{equation}
\begin{matrix}\rmi k_y H_z-\partial_z H_y=-\rmi k_0(\hat\varepsilon_j\vec E)_x, & \rmi k_y E_z-\partial_z E_y=\rmi k_0(\hat\mu_j\vec H)_x, \\
\partial_z H_x-\rmi k_x H_z=-\rmi k_0(\hat\varepsilon_j\vec E)_y, & \partial_z E_x-\rmi k_x E_z=\rmi k_0(\hat\mu_j\vec H)_y, \\
\rmi k_x H_y-\rmi k_y H_x=-\rmi k_0(\hat\varepsilon_j\vec E)_z, &
\rmi k_x E_y-\rmi k_y E_x=\rmi k_0(\hat\mu_j\vec H)_z,
\end{matrix}
\label{eq:maxwelleq}
\end{equation}
where $\partial_z=\partial/\partial z$; $k_x=k_0\sin \theta_0 \cos
\varphi_0$ and $k_y=k_0 \sin \theta_0 \sin \varphi_0$ are tangential
components of the wavevector $\vec k$; $\theta_0$ and $\varphi_0$
are polar and azimuthal angles, respectively, which define the
electromagnetic wave propagation direction; $k_0=\omega/c$ is the
free-space wavenumber; $j=1,2$; $\hat \varepsilon_1$ and $\hat
\mu_2$ are the tensors with $\varepsilon_1$ and $\mu_2$ on their
main diagonal and zeros elsewhere (i.e., $\hat
\varepsilon_1=\varepsilon_1\hat I$, $\hat \mu_2=\mu_2\hat I$, where
$\hat I$ is the identity tensor). From six components of the
electromagnetic field $\vec E$ and $\vec H$, only four are
independent. Thus the components $E_z$ and $H_z$ can be eliminated
from system (\ref{eq:maxwelleq}), and derived a set of four
first-order linear differential equations related to the transversal
field components inside a layer of the structure
\cite{Tuz_JOpt_2010}. The obtained sets of equations can be
abbreviated by using a matrix notation:
\begin{equation}
\partial_z\vec \Psi(z) = \rmi k_0\mathbf A(z)\vec\Psi(z),~~~~0<z<L.
\label{eq:cauchy}
\end{equation}
In this equation, $\vec \Psi = \{E_x,E_y,H_x,H_y\}'$ is a
four-component column vector (here the prime symbol denotes the
matrix transpose operator), while the $4\times 4$ matrix function
$\mathbf A(z)$ is piecewise uniform  as
\begin{equation}
\mathbf A(z)=\left\{\begin{array}{c}
\mathbf A_1,~~~~~0<z<d_1, \\
\mathbf A_2,~~~~~d_1<z<L,
\end{array}\right.
\label{eq:piecewise}
\end{equation}
where the matrices $\mathbf A_1$ and $\mathbf A_2$ correspond to
ferrite and semiconductor layers, respectively.

Since the vector $\vec \Psi$ is known in the plane $z=0$, equation
(\ref{eq:cauchy}) is related to the Cauchy problem
\cite{Jakubovich_book_1975} whose solution is straightforward,
because the matrix $\mathbf A(z)$ is piecewise uniform. Thus, the
field components referred to boundaries of the double-layer period
of the structure are related as\footnote{The series $\exp(\mathbf X)
= \mathbf I +\Sigma_{m=1}^{\infty}\frac{1}{m!}\mathbf X^m$ converges
for square matrices $\mathbf X$, i.e. function $\exp(\mathbf X)$ is
defined for all square matrices \cite{Jakubovich_book_1975}.}
\begin{equation}
\begin{split}
\vec \Psi(L) = \mathbf M_2\vec \Psi(d_1)=\mathbf M_2\mathbf M_1 \vec
\Psi (0)=\mathfrak{M}\vec \Psi (0) =\exp[\rmi k_0\mathbf
A_2d_2]\exp[\rmi k_0\mathbf A_1d_1] \vec\Psi(0),
\end{split}
\label{eq:solution1}
\end{equation}
where $\mathbf M_j$ and $\mathfrak{M}$ are the transfer matrices of
the corresponding layer ($j=1,~2$) and the period, respectively.

Suppose that $\gamma_j$ is the eigenvalue of the corresponding
matrix $k_0\mathbf A_j$ $(\det[k_0 \mathbf A_j-\gamma_j\mathbf
I]=0)$, and $\mathbf I$ is the $4\times 4$ identity matrix. When
$|\gamma_j|d_j\ll1$ (i.e., both layers in the period are optically
thin), the next long-wave approximation can be used
\cite{Lakhtakia_Optic_2003}
\begin{equation}
\exp[\rmi k_0\mathbf A_2d_2]\exp[\rmi k_0\mathbf A_1 d_1]\simeq
\mathbf I+\rmi k_0\mathbf A_1 d_1 +\rmi k_0\mathbf A_2 d_2.
\label{eq:approx1}
\end{equation}

Let us now consider a single layer of effective permittivity $\hat
\varepsilon_{eff}$, effective permeability $\hat \mu_{eff}$ and
thickness $L$. Quantity $\mathbf A_{eff}$ can be defined in a way
similar to (\ref{eq:solution1}):
\begin{equation}
\vec \Psi(L) = \mathfrak{M}_{eff}\vec \Psi (0) =\exp[\rmi k_0\mathbf
A_{eff} L ] \vec\Psi(0). \label{eq:solution2}
\end{equation}

Provided that $\gamma_{eff}$ is the eigenvalue of the matrix
$k_0\mathbf A_{eff}$ $(\det[k_0\mathbf A_{eff}-\gamma_{eff}\mathbf
I]=0)$ and $|\gamma_{eff}|L\ll 1$ (i.e., the entire composite layer
is optically thin as well), the next approximation follows
\begin{equation}
\exp[\rmi k_0\mathbf A_{eff} L]\simeq \mathbf I+\rmi k_0 \mathbf
A_{eff} L. \label{eq:approx2}
\end{equation}

Equations (\ref{eq:approx1}) and (\ref{eq:approx2}) permit us to
establish the following equivalence between bilayer and single
layer:
\begin{equation}
\mathbf A_{eff}=f_1\mathbf A_1+f_2\mathbf A_2,
\label{eq:equivalence}
\end{equation}
where $f_j=d_j/L$, and the elements of the matrices $\mathbf A_1$,
$\mathbf A_2$ and $\mathbf A_{eff}$ one can find in \ref{sec:app_A}.

From equation~(\ref{eq:equivalence}), in the particular case of
coincidence of the directions of both the wave propagation and the
static magnetic field bias ($\theta_0=0$), the following simple
expressions for the effective constitutive parameters of the
homogenized medium can be obtained:
\begin{equation}
\begin{array}{c}
\mu_{eff}^T=f_1\mu_1^T+f_2\mu_2,~~~~~\alpha_{eff}=f_1\alpha,\\
\varepsilon_{eff}^T=f_1\varepsilon_1+f_2\varepsilon_2^T,~~~~~\beta_{eff}=f_2\beta.
\end{array}
\label{eq:effective}
\end{equation}

\begin{figure}[htbp]
\centerline{\includegraphics[width=1.0\columnwidth]{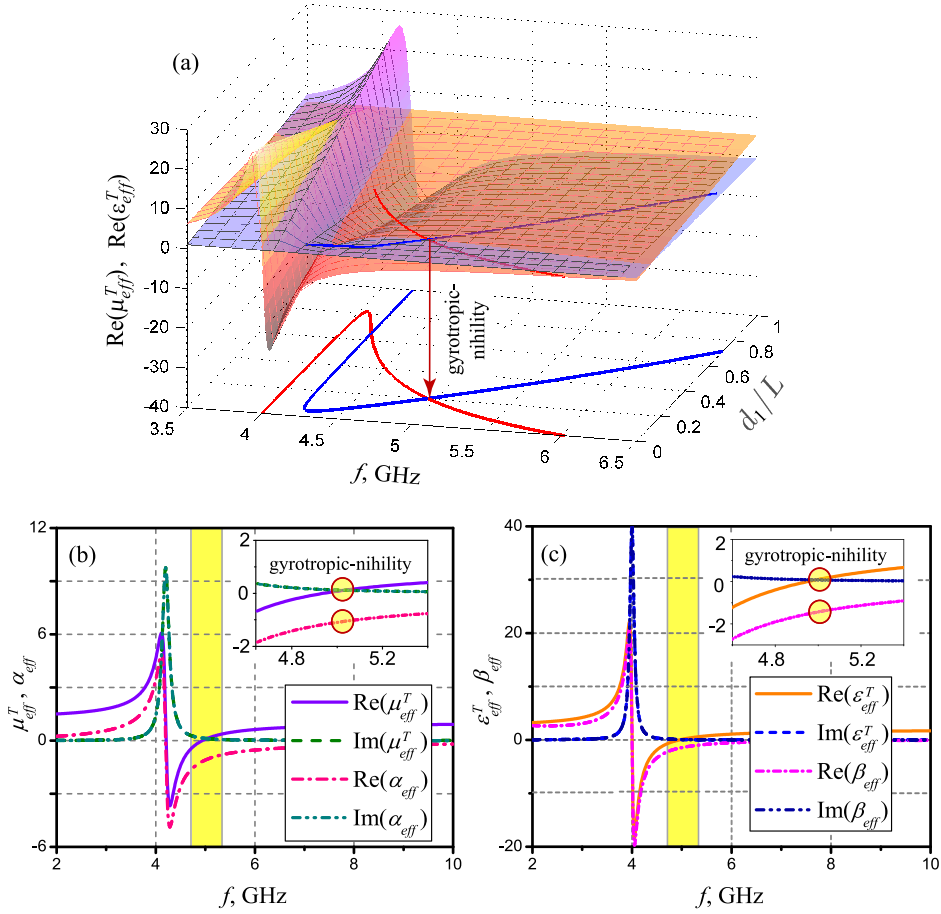}}
\caption{(Color online) (a) Two surfaces depict behaviors of real
parts of effective permeability (purple surface) and  effective
permittivity (orange surface) versus frequency and the ratio of the
layers' thicknesses. The blue and red curves plotted on the surfaces
show the mu-near-zero and epsilon-near-zero states, respectively.
Curves at the bottom are just projections and are given for an
illustrative purpose. Dispersion curves of the tensor components of
(b) effective permeability and (c) effective permittivity at
particular structure parameters ($d_1=0.05$~mm, $d_2=0.2$~mm) for
which the gyrotropic-nihility state exists. Other parameters of the
ferrite and semiconductor layers are the same as in
figure~\ref{fig:fig2}; $L=0.25$~mm.} \label{fig:fig3}
\end{figure}

The effective constitutive parameters calculated according to
equalities~(\ref{eq:effective}) are given in figure~\ref{fig:fig3}.
One can see that there is a frequency range where real parts of both
$\mu_{eff}^T$ and $\varepsilon_{eff}^T$ reach zero. It is
significant that, by special adjusting ferrite and semiconductor
types, the external static magnetic field strength and the ratio of
the layers' thicknesses, it is possible to obtain the condition
where real parts of both $\mu_{eff}^T$ and $\varepsilon_{eff}^T$
simultaneously acquire zero at the same frequency. Note well that at
this frequency, the parameters $\alpha_{eff}$ and $\beta_{eff}$ are
non-zero and the medium losses are relatively small [see,
figures~\ref{fig:fig3}(b) and (c)]. We define exactly this situation
as a gyrotropic-nihility state. In the structure under consideration
such gyrotropic-nihility state appears at a particular frequency
$f_{gn}\approx$4.94~GHz which is distinguished in
figure~\ref{fig:fig3}(a) by an arrow.  It is expectable that in the
general case when the directions of the static magnetic field bias
and the wave propagation do not coincide ($\theta_0 \ne 0$), the
gyrotropic-nihility state can acquire some changes, but nevertheless
it should stay distinguishable. We discuss these changes further
through the eigenwaves and structure's transmittance analysis.

\section{Eigenwaves and gyrotropic-nihility state}

Evidently, the formulation of the eigenvalue problem on the matrix
$\mathbf A_{eff}$ $(\det[\mathbf A_{eff}-\eta\mathbf I]=0)$, whose
coefficients are defined as (\ref{eq:equivalence}), yields us the
characteristic equation on the effective refractive index $\eta$ of
the medium. Thus, we arrive to the following biquadratic equation
with respect to $\eta^2$ \cite{Akhiezer_book_1975,
Rapoport_PhysPlasmas_2010}:
\begin{equation}
A\eta^4+B\eta^2+C=0, \label{eq:characteristic}
\end{equation}
where $A = 1$;
$B=-2(\varepsilon_{eff}^T\mu_{eff}^T+\alpha_{eff}\beta_{eff}) +
(\varepsilon_{eff}^T/\varepsilon_{eff}^L+\mu_{eff}^T/\mu_{eff}^L)(k_x^2+k_y^2)/
k_0^2$ and $C=[(\mu_{eff}^T)^2 - \alpha_{eff}^2- (k_x^2 +
k_y^2)\mu_{eff}^T/k_0^2\varepsilon^L_{eff}]
[(\varepsilon_{eff}^T)^2-\beta_{eff}^2-(k_x^2+k_y^2)\varepsilon_{eff}^T/k_0^2\mu^L_{eff}]$
are known functions of frequency.

One can see that regarding the eigenwaves propagation in an
unbounded gyroelectromagnetic medium, the problem acquires the
cylindrical symmetry which results in the independence of solutions
of biquadratic equation~(\ref{eq:characteristic}) on the azimuthal
angle $\varphi_0$, since there is
$(k_x^2+k_y^2)/k_0^2=\sin^2\theta_0$. Thereby these solutions are
\begin{equation}
\eta_\pm^2(\omega, \theta_0)=\frac{-B\pm\sqrt{B^2-4AC}}{2A} =
\varepsilon_\pm\mu_\pm. \label{eq:solution}
\end{equation}
Obtained root branches describe two effective refractive indexes
$\eta_+$ and $\eta_-$ related to two distinct eigenwaves with
propagation constants $\gamma_+$ and $\gamma_-$
($\gamma_\pm=k_0\eta_\pm$). Those are the ordinary ($+$) and
extraordinary ($-$) waves, respectively, as it is accepted in the
optics community, and in the general case of $\theta_0\ne 0$, these
eigenwaves are elliptically polarized ones. Remarkably that each
eigenwave propagates through its own medium with distinctive
constitutive parameters $\varepsilon_+$, $\mu_+$ and
$\varepsilon_-$, $\mu_-$ for ordinary and extraordinary eigenwave,
respectively.

For lossless medium the electromagnetic waves can propagate through
a medium only if the condition $\eta_\pm^2(\omega, \theta_0)>0$
holds. Nevertheless, equation~(\ref{eq:solution}) does not specify a
sign of the refractive index. So the sign of $\eta$ must be chosen
providing that the energy carrying by the wave goes away from the
source. Here the situation is possible when directions of the
Poynting vector $\vec S=(c/8\pi)\real\{\vec E \times \vec H^*\}$ and
the wavevector $\vec k$ do not coincide and so-called backward
propagation appears. This problem has been discussed before in many
papers related to the double-negative (left-handed) materials
including those addressed on the circularly birefringent
(gyrotropic) media (e.g., see \cite{Lindell_MOP_2001,
Rapoport_PhysPlasmas_2010, Veselago_SovPhysUsp_1968,
Ivanov_JMMM_2006, Chern_JApplPhys_2013}), and so we omit the details
here. Howbeit, in our case the medium losses should be taken into
consideration,  and the signs of the real and imaginary parts of the
complex refractive index $\eta_\pm=\eta_\pm'+\rmi\eta_\pm''$ can be
determined from the solution of the next equation
\cite{Sementsov_PSS_2012}
\begin{equation}
\left(\eta_\pm'+\rmi\eta_\pm''\right)^2=\left(\varepsilon_\pm'+
\rmi\varepsilon_\pm''\right)\left(\mu_\pm'+\rmi\mu_\pm''\right),
\label{eq:refractive}
\end{equation}
which is further reduced to the set of two equations obtained by
deriving the real and imaginary parts from (\ref{eq:refractive})
\begin{equation}
(\eta_\pm')^2-(\eta_\pm'')^2=\varepsilon_\pm'\mu_\pm'-\varepsilon_\pm''\mu_\pm'',~~~
2\eta_\pm'\eta_\pm''=\varepsilon_\pm'\mu_\pm''+\varepsilon_\pm''\mu_\pm'.
\label{eq:refractivetwo}
\end{equation}

Thus, in order to ensure an electromagnetic wave damping as the wave
propagates through the medium, the imaginary parts of permittivity,
permeability and refractive index all must be positive quantities
($\varepsilon_\pm''>0$, $\mu_\pm''>0$, and $\eta_\pm''>0$). From
these conditions it follows that according to the second equation in
(\ref{eq:refractivetwo}) the sign of $\eta_\pm'$ is defined by the
signs and absolute values of $\varepsilon_\pm'$ and $\mu_\pm'$, and
in the particular case of the double-negative medium
($\varepsilon_\pm'<0$, $\mu_\pm'< 0$) the real part of refractive
index $\eta_\pm'$ must be  determined as a negative quantity
\cite{Veselago_SovPhysUsp_1968}.

\begin{figure}[htbp]
\centerline{\includegraphics[width=1.0\columnwidth]{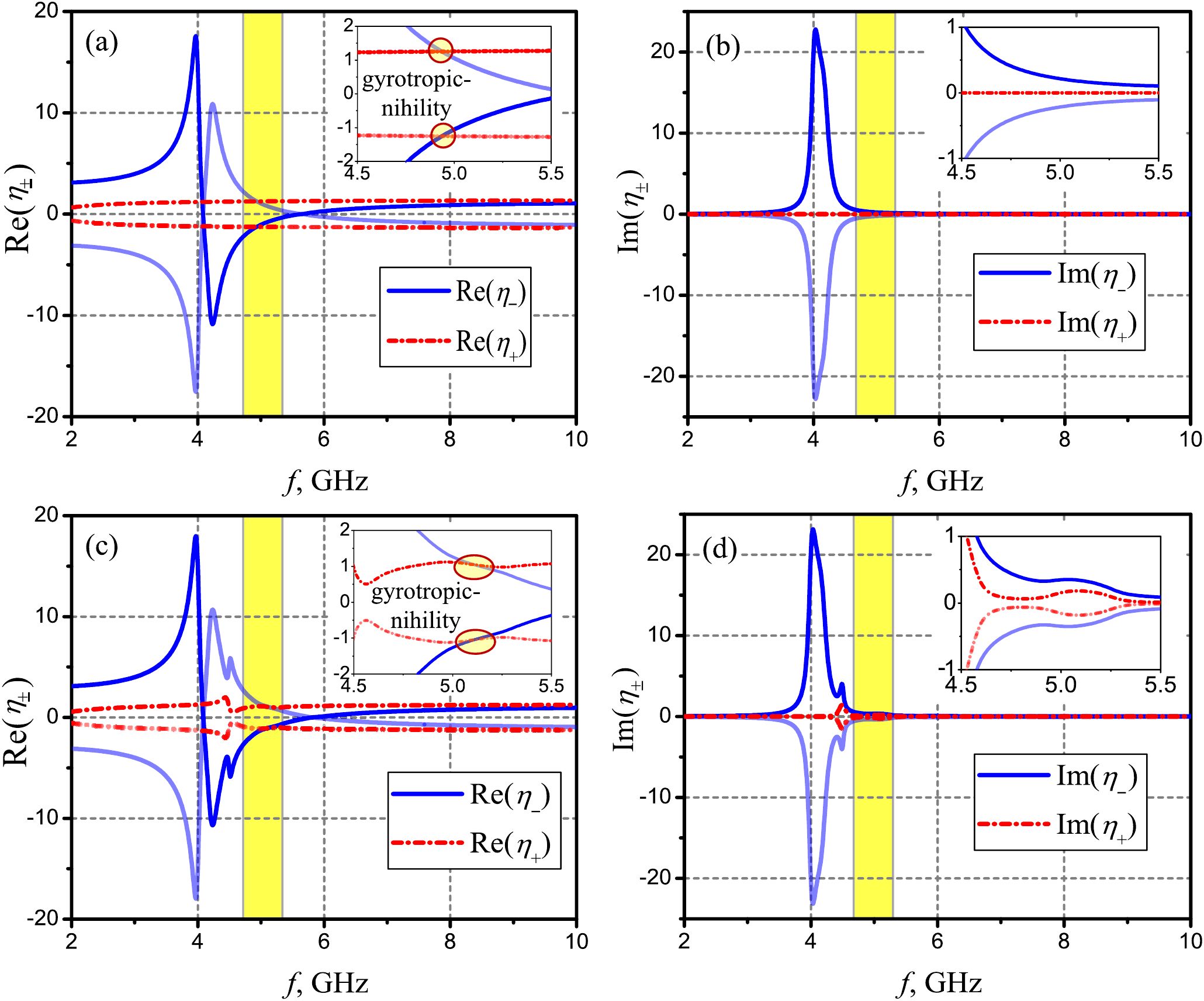}}
\caption{(Color online) Dispersion curves of the refractive indexes
$\eta_\pm$ for (a), (b) $\theta_0=0\deg$ and (c), (d)
$\theta_0=25\deg$. A pair of $\eta_+$ and $\eta_-$ is distinguished
in each plot with bright lines for which conditions $\eta_+''\ge 0$
and $\eta_-''\ge 0$ hold. Curves for which conditions $\eta_+''< 0$
and $\eta_-''< 0$ hold are plotted with pale lines. Parameters of
the ferrite and semiconductor layers are the same as in
figure~\ref{fig:fig2}; $d_1=0.05$~mm, $d_2=0.2$~mm.}
\label{fig:fig4}
\end{figure}

Keeping this in mind, the root branches (\ref{eq:solution}) of
equation~(\ref{eq:characteristic}) are properly chosen and plotted
in figure~\ref{fig:fig4} for two different values of the polar angle
$\theta_0$. Each plot consists of four dispersion curves, from which
we distinguish a pair of effective refractive indexes $\eta_+$ and
$\eta_-$ having positive imaginary parts [see, figures 4(b) and
(c)]. It can be seen that the curves within this pair demonstrate
drastically distinctive features. While the dispersion curve of the
refractive index $\eta_-$ related to the extraordinary eigenwave
undergos a small monotonic increasing with $\eta_-'>1$ and
$\eta_-''\approx 0$, the dispersion curve of the refractive index
$\eta_+$ related to the ordinary eigenwave experiences considerable
changing in which $\eta_+'$ has consistently gone through negative
values to positive ones and $\eta_+''$ acquires some maximum. Thus,
based on characteristics of $\eta_+$, the whole frequency range  of
interest can be divided into three specific bands. The first
frequency band is located between 2~GHz and 4~GHz where  $\eta_+'$
is a positive quantity and $\eta_+''$ is very significant. In the
second band from 4~GHz to 6~GHz, $\eta_+'$  is a negative value
which eventually acquires a transition to positive one on the band
boundaries while $\eta_+''$ decreases as the frequency rises. And
finally, in the third frequency band which starts from 6~GHz,
$\eta_+$ and $\eta_-$ become comparable quantities. Besides, one can
see that in the second frequency band there is a particular
frequency $f_{gn}\approx 4.94$~GHz, where the condition
$\eta_-'=-\eta_+'$ holds which is related to the gyrotropic-nihility
state. The latter situation is marked out in the inset of
figure~\ref{fig:fig4}(a) with circles. Before proceeding to this
particular state, we should note that when $\theta_0\ne 0$ all the
mentioned frequency bands can be still readily distinguished.
However the condition $\eta_-'=-\eta_+'$ of the gyrotropic-nihility
state shifts toward higher frequencies and begins to hold in a
certain frequency band rather than a particular frequency. Exactly
this peculiarity is marked out in the inset of
figure~\ref{fig:fig4}(c) with ellipses.

Fortunately in the case when the electromagnetic waves propagation
is parallel to the magnetic field bias ($\theta_0=0$) equation
(\ref{eq:characteristic}) has simple analytical solutions
\begin{equation}
\eta_\pm^2(\omega,
0)=(\mu_{eff}^T\pm\alpha_{eff})(\varepsilon_{eff}^T\pm\beta_{eff}),
\label{eq:analitical}
\end{equation}
that allows us to analyze in more details the conditions at which
the gyrotropic-nihility state occurs. At the same time, we recall
that when $\eta_\pm'(\omega, 0) > 0$, $\eta_+$ and $\eta_-$ in
equation (\ref{eq:analitical}) are related to two eigenwaves which
propagate along the positive direction of the $z$-axis with right-
and left-circular polarizations, respectively\footnote{We use here
the optical community convention on the handedness of the circular
polarization. According to this convention right-circularly
polarized light is defined as a clockwise rotation of the electric
vector when the observer is looking against the direction the wave
is traveling.}.

From figures \ref{fig:fig3}(b) and (c) one can conclude that in the
whole considered frequency band the imaginary parts of diagonal and
off-diagonal components of both effective permeability and effective
permittivity tensors are close values
($(\mu_{eff}^T)''\approx\alpha_{eff}''$,
$(\varepsilon_{eff}^T)''\approx\beta_{eff}''$), while the real parts
of diagonal elements of both tensors are greater than the real parts
of the corresponding gyrotropic parameters
($(\mu_{eff}^T)'>\alpha_{eff}'$,
$(\varepsilon_{eff}^T)'>\beta_{eff}'$). Since all imaginary parts of
the constitutive parameters must be positive quantities, the sign
and absolute values of their real parts obviously define the sign of
the refractive indexes related to the ordinary and extraordinary
eigenwaves.

Thus, in accordance with the mentioned above characteristics of the
complex effective constitutive parameters, it is appreciated that in
the whole considered frequency band the refractive index $\eta_-$
related to the extraordinary eigenwave is a positive quantity with
the vanishingly small imaginary part. At once, the refractive index
$\eta_+$ related to the ordinary eigenwave can be either positive or
negative quantity in the corresponding frequency bands. In
particular it becomes a negative quantity in the frequency bands
where the next conditions hold: (i) $(\mu_{eff}^T)' < 0$,
$(\varepsilon_{eff}^T)'<0$; (ii) $\alpha_{eff}'<0$, $\beta_{eff}'<0$
and $|(\mu_{eff}^T)'|<|\alpha_{eff}'|$,
$|(\varepsilon_{eff}^T)'|<|\beta_{eff}'|$; (iii) $(\mu_{eff}^T)' <
0$, $|(\varepsilon_{eff}^T)'|<|\beta_{eff}'|$ or
$|(\mu_{eff}^T)'|<|\alpha_{eff}'|$, $(\varepsilon_{eff}^T)'<0$, and
it is evident that in all these bands the imaginary part of $\eta_+$
cannot be zero. From our estimations the figure of merit
$\Phi_\pm=|\eta_\pm'|/\eta_\pm''$ is the order of tens and hundreds
for ordinary and extraordinary eigenwaves, respectively. It means
that there is a different damping of the ordinary and extraordinary
eigenwaves as they propagate through the medium, which is a known
manifestation of the circular dichroism.

\begin{figure}[htbp]
\centerline{\includegraphics[width=1.0\columnwidth]{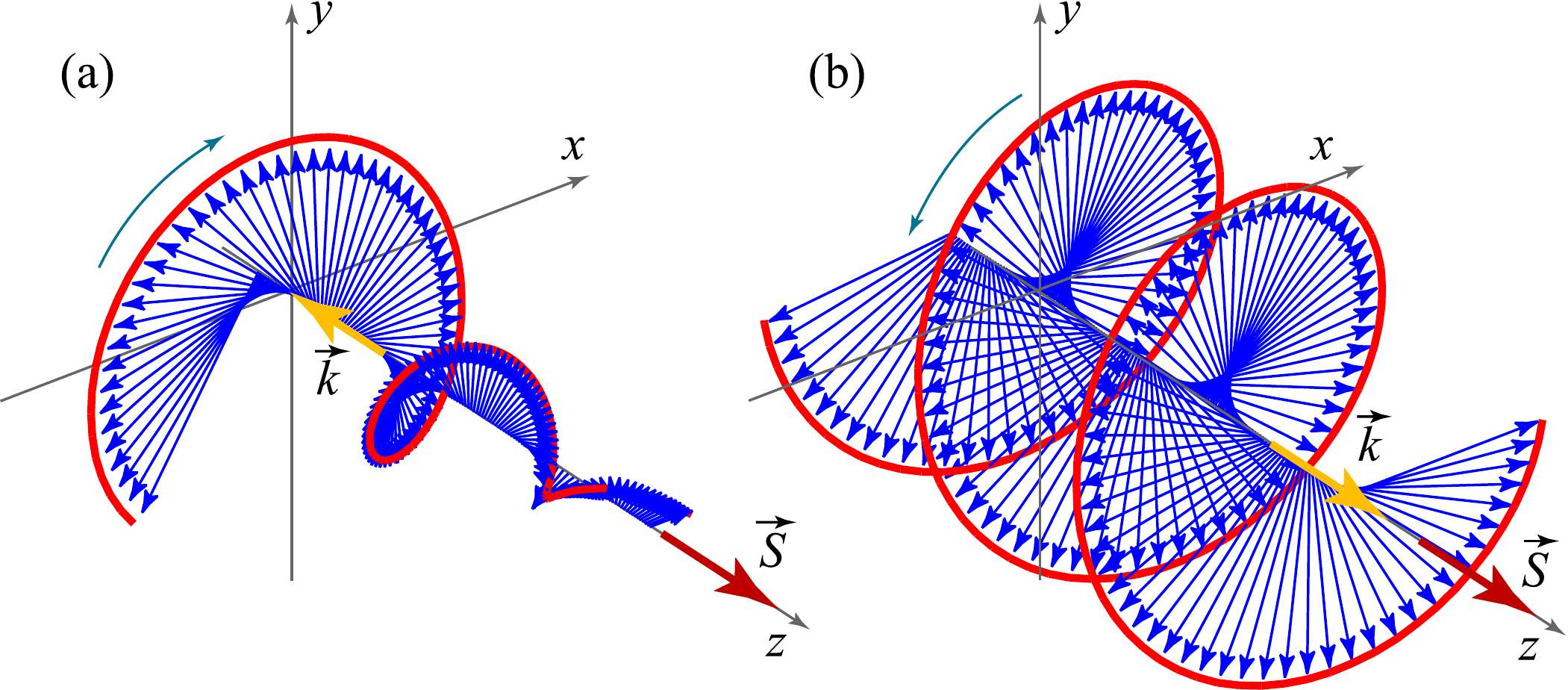}}
\caption{(Color online) Instantaneous electric field distribution
for (a) ordinary eigenwave and (b) extraordinary eigenwave at the
particular frequency of the gyrotropic-nihility state $f_{gn}=4.94$.
Parameters of the ferrite and semiconductor layers are the same as
in figure~\ref{fig:fig2}; $\theta_0=0\deg$, $d_1=0.05$~mm,
$d_2=0.2$~mm.} \label{fig:fig5}
\end{figure}

Remarkably, from this set of conditions, the second one is
unambiguously satisfied at the particular frequency of the
gyrotropic-nihility state $f_{gn}$, where $(\mu_{eff}^T)'$ and
$(\varepsilon_{eff}^T)'$ simultaneously appear to be close to zero.
In this case the imaginary parts of $\mu_{eff}^T$, $\alpha_{eff}$
and $\varepsilon_{eff}^T$, $\beta_{eff}$ are relatively small, so we
can write $|\varepsilon_\pm \mu_\pm| \approx
|\alpha_{eff}'\beta_{eff}'|$, which leads to the following
approximate equalities related to the eigenwaves propagation
constants
\begin{equation}
\gamma_-\approx -\gamma_+\approx
k_0\sqrt{|\alpha_{eff}'\beta_{eff}'|}. \label{eq:analitprop}
\end{equation}
Thus, the propagation constants of the ordinary and extraordinary
eigenwaves are equal in the magnitude but opposite in sign to each
other, and thus a backward propagation appears for the ordinary
eigenwave while for the extraordinary eigenwave it is a forward one.
So, it turns out that at the frequency of the gyrotropic-nihility
state, both ordinary $\gamma_+$ and extraordinary $\gamma_-$
eigenwaves appear to be left-circularly polarized. This is because
the wavevector $\vec k$ of the ordinary eigenwave now reverses its
direction and the handedness changes, accordingly, from left to
right (see, figure~\ref{fig:fig5}).

It is particularly remarkable that at the gyrotropic-nihility state,
the effective gyrotropic parameters $\alpha_{eff}$ and $\beta_{eff}$
are close in value to each other and their real parts simultaneously
approach unity which can be clearly seen in figure~\ref{fig:fig3}.
It leads to the fact that the studied medium appears to be matched
to free space with both the refractive index and the wave impedance.
Directly at the gyrotropic-nihility frequency $f_{gn}$, the
normalized impedances related to the ordinary and extraordinary
waves are almost indistinguishable and are both positive quantities
because when calculating the normalized impedance it is need to take
the root branch as $Z_\pm=\sqrt{\mu^\pm/\varepsilon^\pm}>0$
\cite{Veselago_SovPhysUsp_1968}
\begin{equation}
Z_+\approx Z_- \approx\sqrt{\frac{|\alpha_{eff}'|}{|\beta_{eff}'|}},
\label{eq:impedance}
\end{equation}
Therefore it turns out that this simultaneous matching of both the
refractive index and the wave impedance to free space should
inevitably result in the reflectionless interaction of
electromagnetic waves when they impinge on the studied structure
having a finite number of periods.

\section{Wave transmission through a gyrotropic-nihility layer}

In order to find the transmittance and reflectance of the studied
structure, we use the results of \cite{Tuz_JOpt_2010} and write the
solution of equation (\ref{eq:cauchy}) in the form
\begin{equation}
\vec \Psi(0) = (\mathfrak{M}^N)^{-1}\vec \Psi (NL) =  \mathfrak{T}
\vec \Psi (NL), \label{eq:solution2}
\end{equation}
where the field vector  $\vec \Psi$ at the input and output
structure's surfaces consists of the incident, reflected and
transmitted wave contributions as
\begin{equation}
\vec \Psi(0) =\vec \Psi_{inc}+\vec \Psi_{ref},~~~ \vec \Psi
(NL)=\vec \Psi_{tr}. \label{eq:fields_irt}
\end{equation}
The vectors $\vec \Psi_{inc}$,  $\vec \Psi_{ref}$ and $\vec
\Psi_{tr}$ are composed from tangential  components of the
electromagnetic field, and in turn these components are determined
by their complex amplitudes. In the general case, the numerical
solution of the Cauchy problem (\ref{eq:cauchy}) for particular
structure's layers results in the matrices $\mathbf M_1$ and
$\mathbf M_2$, and then subsequent rising of their product
$\mathfrak{M}=\mathbf M_2\mathbf M_1$ to the power $N$ allows us to
find both the coefficients of the transfer matrix  $\mathfrak{T}$
and  the complex amplitudes of the transmitted and reflected fields
(we refer the reader to Ref.~\cite{Tuz_JOpt_2010} here for further
details on the calculation procedure). The ratios between the
amplitudes of the transmitted field and the incident field, and
between the amplitudes of the reflected field and the incident field
establish the complex  transmission and reflection coefficients,
respectively, which can be calculated as functions of the frequency
and angle of wave incidence, $T(\omega, \theta_0)$ and $R(\omega,
\theta_0)$. Accordingly, the quantities $|T|^2$, $|R|^2$ and
$W=1-|T|^2-|R|^2$ are defined as the transmittance, reflectance and
absorption coefficient.

\begin{figure}[htbp]
\centerline{\includegraphics[width=1.0\columnwidth]{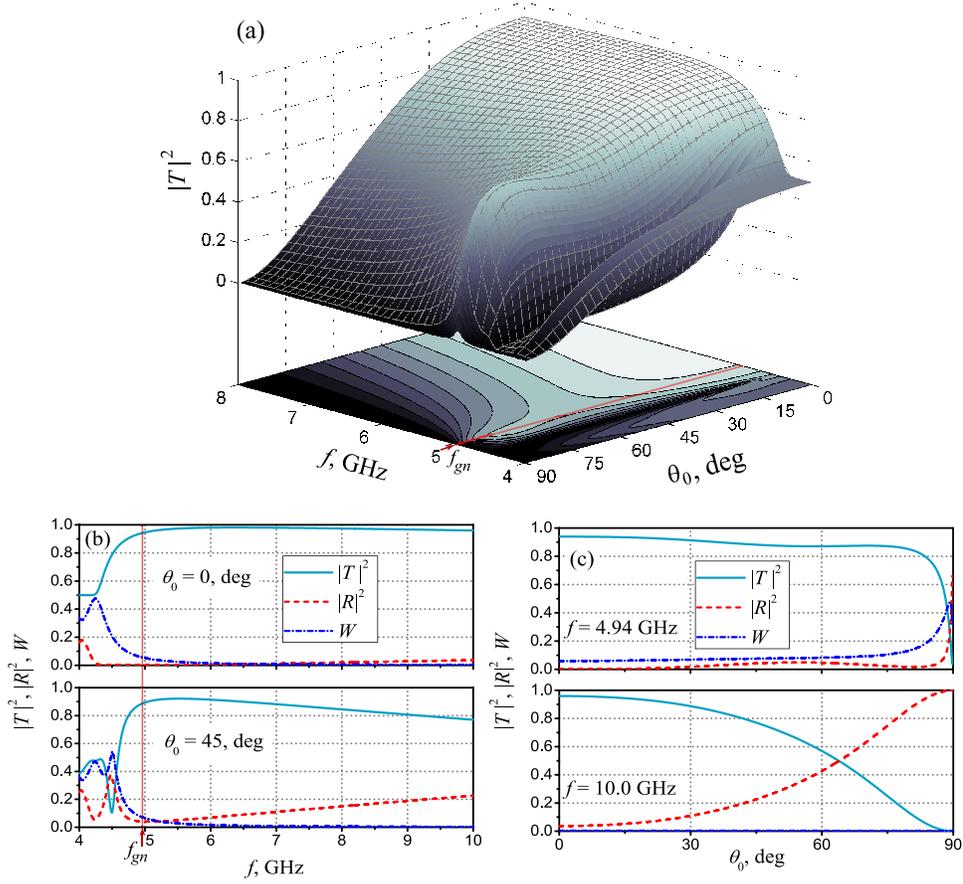}}
\caption{(Color online) (a) Transmittance as a function of the
frequency and the polar angle of incidence of the plane
monochromatic wave for a finely-stratified structure with $N=10$
periods which corresponds to the equivalent gyroelectromagnetic
layer with thickness $NL=2.5$~mm. (b), (c) The frequency and angular
dependencies of the transmittance, reflectance and absorption
coefficient for the same structure. Parameters of the ferrite and
semiconductor layers are the same as in figure~\ref{fig:fig2};
$d_1=0.05$~mm, $d_2=0.2$~mm.} \label{fig:fig6}
\end{figure}

In figure~\ref{fig:fig6}(a) the transmittance calculated as a
function of the frequency and angle of incidence is shown in the
from of a surface plot. One can see that this surface is quite
smooth on the distance from the ferromagnetic and plasma resonances.
At once, in the frequency band of interest (4.5~GHz~--~5.5~GHz)
there is a specific ridge on the surface where the transmittance
reaches high values. This ridge runs through almost the entire range
of angles, and maximum of  the transmittance appears near the
frequency of the gyrotropic-nihility state $f_{gn}=4.94$~GHz, which
is distinguished on the bottom contour by an arrow. Such high
transmittance obviously appears due to the mentioned peculiarities
of the refractive index and the wave impedance that are both matched
to free space.

This feature is also confirmed by the curves plotted in the bottom
planes of figure~\ref{fig:fig6}, where the transmittance,
reflectance and absorption coefficient are presented for two
different values of the frequency and polar angle. Here again, the
frequency of the gyrotropic-nihility state is marked by an arrow.
From curves plotted in figure~\ref{fig:fig6}(b) one can conclude
that despite the fact that the angle of incidence rises the minimum
of reflectance remains to be nearly the frequency of the
gyrotropic-nihility state wherein a certain absorption in the medium
exists.

Besides, in figure~\ref{fig:fig6}(c) the first frequency is chosen
at the gyrotropic-nihility state while the second one is selected to
be far from the frequencies of the gyrotropic-nihility state and the
ferromagnetic and plasma resonances. At the frequency of $f =
10$~GHz, the curves have typical form where the transmittance
monotonically decreases and the reflectance monotonically increases
as the angle of incidence rises. On the other hand, at the frequency
of the gyrotropic-nihility state, the curves of the transmittance
and reflectance are different drastically from those of the
discussed case. Thus, the level of the transmittance/reflectance
remains to be invariable almost down to the glancing angles. At the
same time, the reflectance is small down to the glancing angles
because at this frequency the medium is matched to free space.

\section{Conclusions}

To conclude, in this paper we study characteristics of the
gyrotropic-nihility state in a finely-stratified
ferrite-semiconductor structure which is under an action of an
external static magnetic field applied in the Faraday configuration.
In the long-wavelength limit, when the structure's layers as well as
its period are optically thin, with an assistance of the effective
medium theory,  the studied structure is approximated as a uniform
gyroelectromagnetic medium defined with effective permittivity and
effective permeability tensors. In general, the investigations of
the eigenwaves propagation in such gyroelectromagnetic medium were
carried out on the basis of numerical calculations. At the same
time, in the case, when directions of the electromagnetic waves
propagation and the static magnetic field bias are coincident, the
components of effective permittivity and effective permeability
tensors, effective refractive indexes and normalized wave impedances
are obtained analytically.

The gyrotropic-nihility phenomenon is considered as some
extreme-parameter state that appears  in a small region near
singular points of dispersion curves where real parts of the
diagonal components of both complex effective permittivity and
complex effective permeability tensors simultaneously make
transitions from negative to positive values while the off-diagonal
components of the corresponding tensors remain to be non-zero
quantities. On the basis of these constitutive parameters the
peculiarities of the ordinary and extraordinary eigenwaves
propagation are studied and the possibility of achieving a
double-negative condition at a particular frequency of the
gyrotropic-nihility state is predicted. In particular, it turns out
that the propagation constants of the ordinary and extraordinary
eigenwaves are equal in the magnitude but opposite in sign to each
other, and thus a backward propagation appears for the ordinary
eigenwave while for the extraordinary eigenwave it is a forward one.
Therefore, at the frequency of the gyrotropic-nihility state, both
ordinary and extraordinary eigenwaves appear to be left-circularly
polarized because the wavevector of the ordinary eigenwave reverses
its direction and the handedness changes, accordingly, from left to
right.

The frequency and angular dependencies of the transmittance,
reflectance and absorption coefficient are presented. It turns out
that near the gyrotropic-nihility state the studied structure is
matched to free space with both the refractive index and the wave
impedance which results in its high transmittance almost in the
entire range of angles of the electromagnetic waves incidence. We
believe that this outcome can be of great interest, particularly, in
the problem of transformation optics.

\appendix
\section{}
\label{sec:app_A}

The matrix $\mathbf A$ in equation~(\ref{eq:cauchy}) can be written
in the $2\times 2$ block representation \cite{Tuz_PIERB_2011}
\begin{equation}
\mathbf A= \begin{pmatrix}
\mathbf 0 & \mathbf A^+ \\
\mathbf A^- & \mathbf 0 \\
\end{pmatrix},
\label{eq:matrix_A}
\end{equation}
where $\mathbf 0$ is the $2\times 2$ matrix with all its entries
being zero, and for the ferrite ($\mathbf A_1$), semiconductor
($\mathbf A_2$) and entire composite ($\mathbf A_{eff}$) layers
corresponding matrices $\mathbf A^\pm$ are \cite{Tuz_PIERB_2012}:
\begin{equation}
\mathbf A^+_1= \begin{pmatrix}
k_x k_y/k_0^2\varepsilon_1+i\alpha & \mu_1^T-k_x^2/k_0^2\varepsilon_1  \\
-\mu_1^T+k_y^2/k_0^2\varepsilon_1 & -k_x k_y/k_0^2\varepsilon_1+i\alpha \\
\end{pmatrix},
\label{eq:matrix1}
\end{equation}

\begin{equation}
\mathbf A^-_1= \begin{pmatrix}
-k_x k_y/k_0^2\mu_1^L & -\varepsilon_1+k_x^2/k_0^2\mu_1^L \\
\varepsilon_1-k_y^2/k_0^2\mu_1^L & k_x k_y/k_0^2\mu_1^L \\
\end{pmatrix},
\label{eq:matrix2}
\end{equation}

\begin{equation}
\mathbf A^+_2= \begin{pmatrix}
k_x k_y/k_0^2\varepsilon_2^L & \mu_2-k_x^2/k_0^2\varepsilon_2^L \\
-\mu_2+k_y^2/k_0^2\varepsilon_2^L & -k_x k_y/k_0^2\varepsilon_2^L \\
\end{pmatrix},
\label{eq:matrix3}
\end{equation}

\begin{equation}
\mathbf A^-_2= \begin{pmatrix}
-k_x k_y/k_0^2\mu_2-i\beta & -\varepsilon_2^T+k_x^2/k_0^2\mu_2 \\
\varepsilon_2^T-k_y^2/k_0^2\mu_2 & k_x k_y/k_0^2\mu_2-i\beta \\
\end{pmatrix},
\label{eq:matrix4}
\end{equation}

\begin{equation}
\mathbf A^+_{eff}= \begin{pmatrix}k_x k_y/k_0^2\varepsilon_{eff}^L+i\alpha_{eff} & \mu_e^T-k_x^2/k_0^2\varepsilon_e^L \\
-\mu_{eff}^T+k_y^2/k_0^2\varepsilon_{eff}^L & -k_x k_y/k_0^2\varepsilon_{eff}^L+i\alpha_{eff} \\
\end{pmatrix},
\label{eq:matrix5}
\end{equation}

\begin{equation}
\mathbf A^-_{eff}= \begin{pmatrix}-k_x k_y/k_0^2\mu_{eff}^L-i\beta_{eff} & -\varepsilon_{eff}^T+k_x^2/k_0^2\mu_{eff}^L \\
\varepsilon_{eff}^T-k_y^2/k_0^2\mu_{eff}^L & k_x k_y/k_0^2\mu_{eff}^L-i\beta_{eff} \\
\end{pmatrix}.
\label{eq:matrix6}
\end{equation}

\ack

This work was supported by National Academy of Sciences of Ukraine
with Program `Nanotechnologies and Nanomaterials,' Project
no.~1.1.3.17.


\section*{References}


\bibliographystyle{vancouver}

\bibliography{Tuz_nihility}


\end{document}